\documentclass[twocolumn,english,aps,pra]{revtex4-1}
\usepackage[T1]{fontenc}
\usepackage[latin9]{luainputenc}
\usepackage{amsmath}
\usepackage{amssymb}
\usepackage{graphicx}
\usepackage{float}
\usepackage{xcolor}
\usepackage{soul}
\setstcolor{red}

\makeatletter
\begin{document}
\title{$\mathcal{PT}$-symmetry effects in measurement-based quantum thermal machines}

\author{Jonas F. G. Santos}
\email{jonassantos@ufgd.edu.br}
\affiliation{Faculdade de Ci\^{e}ncias Exatas e Tecnologia, Universidade Federal da Grande Dourados, Caixa Postal 364, Dourados, CEP 79804-970, MS, Brazil}

\affiliation{Centro de Ci\^{e}ncias Naturais e Humanas, Universidade Federal do ABC, Avenida dos Estados 5001, 09210-580, Santo Andr\'{e}, S\~{a}o Paulo, Brazil}

\author{Pritam Chattopadhyay}
\email{pritam.cphys@gmail.com}
\affiliation{AMOS and Department of Chemical and Biological Physics, Weizmann Institute of Science, Rehovot 7610001, Israel}

\begin{abstract}
Measurement-based quantum thermal machines are fascinating models of thermodynamic cycles where measurement protocols play an important role in the performance and functioning of the cycle. Despite theoretical advances, interesting experimental implementations have been reported. Here we move a step further by considering a measurement-based quantum thermal machine model where the only thermal bath is structured with $\mathcal{PT}$-symmetric non-Hermitian Hamiltonians.  Theoretical results indicate that $\mathcal{PT}$-symmetric effects and measurement protocols are related along the cycle. Furthermore, tuning the parameters suitably it is possible to improve the absorbed heat and the extracted work, operating in the Otto limit for the efficiency, provided we consider a quasi-static cycle. Our model also allows switching the configuration of the cycle, engine, or refrigerator, depending on the strength of the measurement protocol.

\end{abstract}

\maketitle

\section{Introduction}

The development of classical thermodynamics is a solid ground for the theoretical and practical study of thermal machines. For example, using the association between one thermal engine and one refrigerator it is possible to show the second law of classical thermodynamics and that no design of classical machines is able to overcome the Carnot bounds~\cite{Callenbook}. As the systems involved in the fabrication of machines get miniaturized, the working substance or the refrigerant system being the spin (or spins) of an atom (atoms)~\cite{Peterson2018, Henao2018, Lisboa2022} or the vibrational modes of an ion~\cite{Rossnagel2016, Maslennikov2019}, quantum fluctuations become relevant in the proper understanding of the energy transfer and the entropy production during any cyclic operation~\cite{Esposito2009, Campisi2011}. Quantum thermal machines (QTM)~\cite{Scovil1959, Quan2007, Feldmann2000, Gelbwaser2013, Misra2022, Mukhopadhyay2018, Camati2019, Singh2020,chand1,chand2,Pritam2020} are the grounds for the so-called quantum thermodynamics. By investigating QTM we can study quantum effects such as entanglement and coherence and how they affect the performance of QTM. It is also possible to understand and elaborate quantum fluctuation theorems~\cite{Esposito2009, Campisi2011, Jarzynski2004, Seifert2012} as well as thermodynamic uncertainty relations~\cite{Timpanaro2019, Lee2021, Sacchi2021, Pritam2019, Pritam2021,Pritam2020,Pritam2021a}. In recent years, quantum measurements have also played a crucial role in designing new QTM models, as they can change the state of a system and modify its internal energy~\cite{Yi2017, Elouard2018, Elouard2017}. Measurement protocols are non-unitary operations acting on a quantum system. Due to this aspect, many recent QTM's replace one thermalization process with a measurement protocol and the performance becomes dependent on how the measurement is implemented. 

Protocols involving quantum measurements are vast in the literature~\cite{Wiseman2009}, and they are the basis of quantum mechanics theory. To get an interpretation of the non-ideal measurement, Hendrikx~\cite{Muynck2001} in their work considered Ramsay's experiment to explore the generalized measurement protocols. While in projective measurements the system state is completely collapsed in a particular eigenstate after the measurement, in the so-called weak measurement protocols~\cite{Clerk2010, Svensson2013} the system is only slightly perturbed resulting in many applications in quantum information~\cite{Kim2012}. On the other hand, generalized measurements are designed such that one can control the action on the quantum system and, depending on the parameter strength involved, it is possible to vary between weak and projective measurements. Generalized measurements are well described by the positive operator-valued measurements (POVMs). There is a growing interest in designing QTM where at least one of the steps is performed by quantum measurements, where it plays the role of non-unitary processes, thus relating quantum measurements to entropy production and irreversibility. In particular, recently an experimental implementation of QTM using generalized measurements in nuclear magnetic resonance setup was reported~\cite{Lisboa2022}.

The theory of quantum mechanics supports an interesting generalization well-known as $\mathcal{PT}$-symmetric quantum mechanics, which mathematically means that the property of Hermiticity for the observables is relaxed and replaced by the $\mathcal{PT}$-symmetric conditions, i.e., the operators have to fulfill the conditions of invariance by spatial reflection (parity $\mathcal{P}$) and time reversal $\mathcal{T}$ in order to assure real eigenvalues and thus they can represent physical systems~\cite{Bender1998, Bender2015}.  From the seminal paper by Bender and Boettcher~\cite{Bender1998}, it became evident that this new class of Hamiltonians, so-called $\mathcal{PT}$-symmetric Hamiltonians, could have a huge set of applications in quantum physics, for instance, in fluctuation relations theorems~\cite{Deffner2015, Zeng2017, Gardas2016}, in quantum optics and photonics systems~\cite{Ruter2010, Regensburger2002, Gao2021, Xue2022}. In open quantum systems, Ref.~\cite{Bartlomiej2016} showed that the decoherence dynamics are considerably modified if the system-environment interaction is built by $\mathcal{PT}$-symmetric Hamiltonians~\cite{Duarte2018, Dey2019}. Furthermore, Ref.~\cite{Santos2021} proposed a thermal reservoir model with $\mathcal{PT}$-symmetric Hamiltonians based on a collisional model such that modifying the strength of the $\mathcal{PT}$-symmetry is sufficient to switch the configuration from the engine to refrigerator. Although there exists the study of $\mathcal{PT}$-symmetry properties in quantum thermodynamics, their applications in thermodynamic cycles are still unclear for general cycles besides the standard quantum Otto cycle.

In this work, we extend the study of measurement-based thermal machines models by assuming that the only thermal bath in the cycle is structured through $\mathcal{PT}$-symmetric Hamiltonians, while the working substance is assumed to be described by a Hermitian Hamiltonian. This assumption guarantees that the advantage in the performance comes from the thermal bath or of the measurement protocol. Since the measurement mechanism playing the role of a thermal reservoir introduces a new class of quantum thermal machines, the introduction of a second non-trivial degree of freedom may be interesting to control the energy flow along the cycle and then modify its structure and performance. With this in mind, we propose a measurement-based cycle in which the working substance is a single harmonic oscillator, for instance, the vibration mode of a trapped ion, with one of the non-unitary processes a measurement protocol, whereas the second is the interaction with a thermal reservoir modeled by $\mathcal{PT}$-symmetric Hamiltonians via collisional model. This work is organized as follows. In section \ref{reviewsection} we review the main properties of $\mathcal{PT}$-symmetric quantum mechanics and generalized measurements for continuous variable systems. Section \ref{cycledescription} is dedicated to describe the proposed cycle in detail and discuss its consequences. The conclusion and final remarks are drawn in section \ref{conclusion}.

\section{$\mathcal{PT}$-symmetric quantum mechanics and generalized measurements} \label{reviewsection}

\subsection*{$\mathcal{PT}$-symmetric quantum mechanics}

We briefly revisit the main aspects of the $\mathcal{PT}$-symmetric quantum mechanics. In order to represent physical observables, quantum mechanics imposes that operators have to be Hermitian, $A = A^\dagger$, such that they have a complete set of eigenstates and real spectra. As evidenced in the pioneering work~\cite{Bender1998}, non-Hermitian operators that are simultaneously invariant under parity $\mathcal{P}$ and time reversal $\mathcal{T}$ symmetries also fulfill the conditions to represent physical observables. This property is well-known as $\mathcal{PT}$-symmetry and for a non-Hermitian Hamiltonian $H\left(q_i, p_i\right)$ with $i=1...,N$ and with eigenstates $|\psi\left(t\right)\rangle$, it implies $\left[H\left(q_i, p_i\right), \mathcal{PT}\right] = 0$ as well as $\mathcal{PT}|\psi\left(t\right)\rangle = |\psi\left(t\right)\rangle$. Under this condition, the Hamiltonian is called $\mathcal{PT}$-symmetric and invariant under the transformations 
\begin{eqnarray} \nonumber
    \mathcal{PT}q_i\left(\mathcal{PT}\right)^{-1} &=& - q_i, \\ \nonumber \mathcal{PT}p_i\left(\mathcal{PT}\right)^{-1} &=&  p_i\\ 
    \mathcal{PT}i\left(\mathcal{PT}\right)^{-1} &=& - i.
\end{eqnarray}

Apart from the real spectra of a non-Hermitian Hamiltonian fulfilling the $\mathcal{PT}$-symmetry condition, its connection with its Hermitian partner is realized through the similarity transformation~\cite{Mostafazadeh2004, Fring2016}
\begin{equation}
h\left(q_i, p_i\right) = \eta H\left(q_i, p_i\right)\eta^{-1}\label{simtrans},
\end{equation}
with $h\left(q_i, p_i\right)$ the Hermitian partner of $H\left(q_i, p_i\right)$ and $\eta = \eta\left(q_i, p_i\right)$ is the Dyson map with the property $\eta \eta^{-1} = \mathbb{I}$. $\mathcal{PT}$-symmetric Hamiltonians satisfying Eq. (\ref{simtrans}) together with the Hermitian condition guarantee the quasi-Hermiticity relation, $\Theta H\left(q_i, p_i\right) = H^\dagger\left(q_i, p_i\right) \Theta$, with $\Theta = \eta^\dagger \eta$ the metric operator in order to ensure the probability conservation~\cite{Fring2016, Luiz2020}. For Hermitian Hamiltonians, $\Theta = \mathbb{I}$. Finally, expected values of observables of $\mathcal{PT}$-symmetric non-Hermitian Hamiltonians and their Hermitian partner are linked through the relation $\langle \phi \left(t\right)|O| \phi \rangle =  \langle \psi \left(t\right)|\mathcal{O}| \psi \rangle$, where $\mathcal{O}$ and $O$ are the non-Hermitian and Hermitian partner observables, respectively, and $|\phi \left(t\right) \rangle = \eta^{-1} |\psi \left(t\right) \rangle$~\cite{Luiz2020}. 

\subsection*{Generalized measurements} 

Recent studies concerning measurement protocols in quantum mechanics assume generalized measurement as the most general set of measurements~\cite{Wiseman2009}. They range from the class of projective to weak measurements, depending on their strength. Generalized measurements of such a observable $A$ with eigenvalues $a_\alpha$ are characterized by Hermitian measurement operators $M_\alpha = M_\alpha^\dagger$ such that $\sum_\alpha M_\alpha^2 = \mathbb{I}$~\cite{Yi2017}. The state of the system after the measurement in the non-selective case has the form $\rho^M = \sum_\alpha M_\alpha \rho M_\alpha^\dagger$. 
Quantum thermal machines powered by generalized measurements have been recently reported in~\cite{Lisboa2022, Behzadi2021}. 

Generalized measurements have been used in quantum thermal machine models with qubit playing the role of working substance in Ref.~\cite{Lisboa2022, Yi2017, Behzadi2021}. For continuous variable systems, we can perform Gaussian measurements on the position or momentum operators such that $M_\alpha = \left(2\pi \sigma^{2}\right)^{-1/4} \text{exp}\left[-\left(q - \alpha\right)^2/\left(4\sigma^2\right)\right]$ for the position case, with $\alpha$ the measured position and $\sigma^2$ is the variance of the measurement apparatus and characterizes its precision. Here the measurement operators satisfy the normalization condition $\int d\alpha\, M^2_\alpha = \mathbb{I}$. Note that for $\sigma^2 \rightarrow 0$  we have infinite precision in the measurement which is in agreement with a projective measurement protocol. Also, these measurement operators correspond to the homodyne probability density associated with a measurement of the position~\cite{Serafinibook}. It is important to stress the non-unitary aspect of this generalized measurement, such that after the measurement protocol the system in general absorbs or releases an amount of heat, whenever the Hamiltonian and the measurement operators do not commute. It is interpreted as heat because the Hamiltonian is kept fixed during the process.

\section{$\mathcal{PT}$-symmetry in a based-measure quantum heat engine} \label{cycledescription}

\begin{figure}
\centering
\includegraphics[scale=0.70]{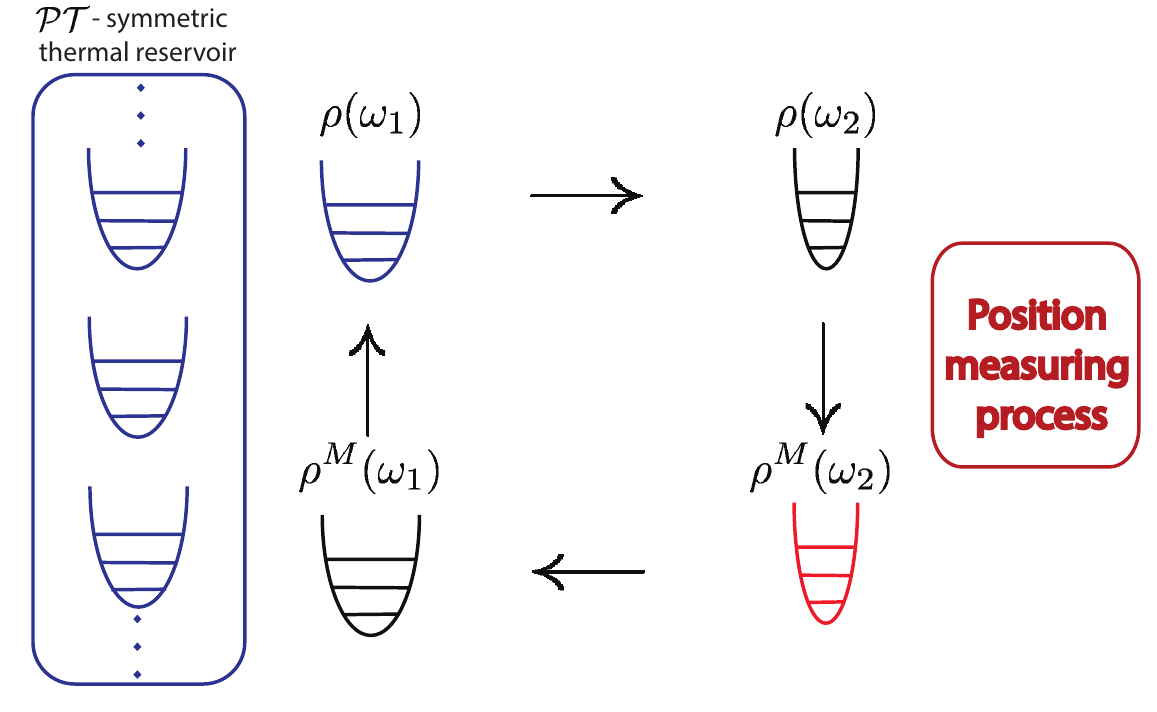}
\includegraphics[scale=0.55]{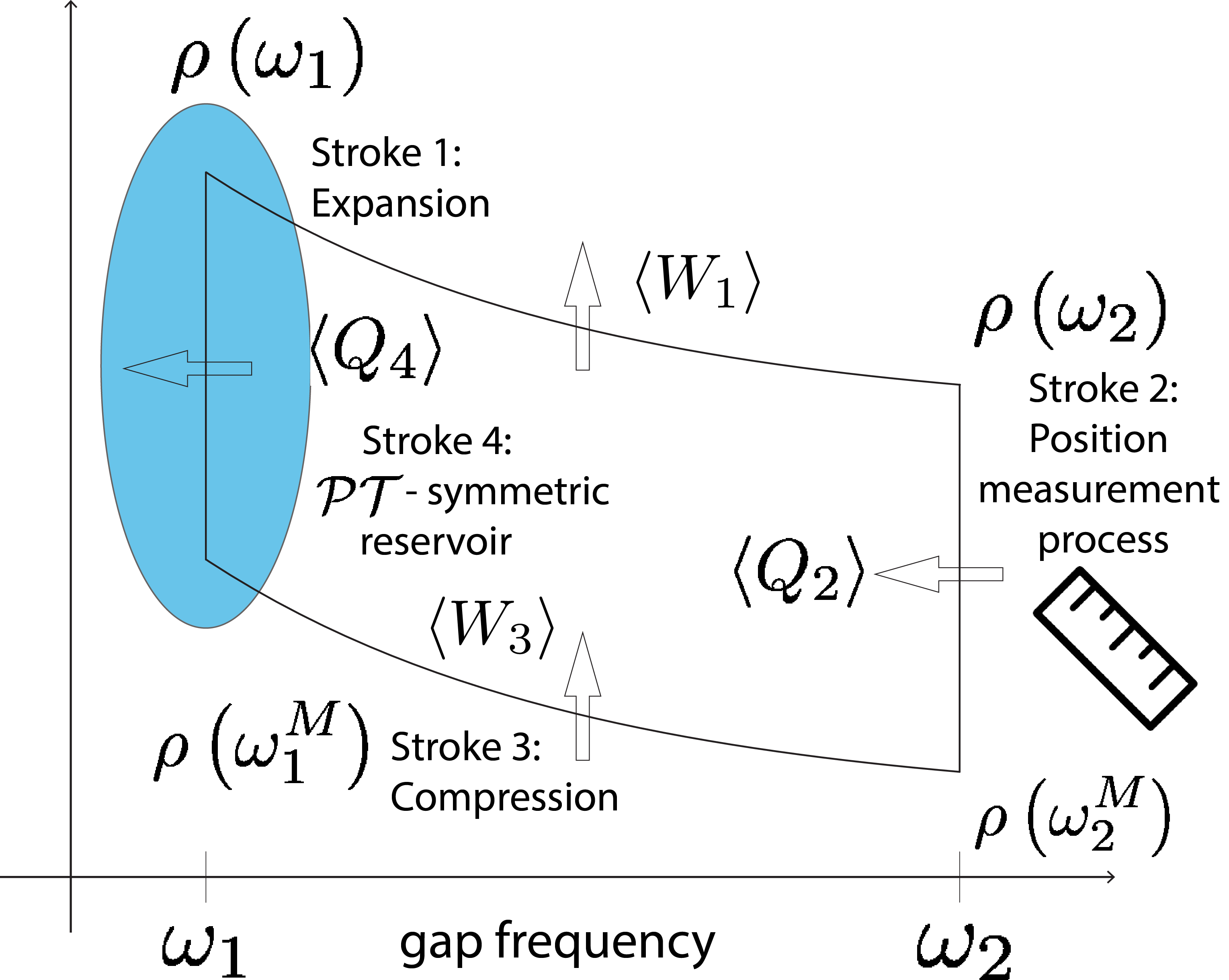}

\caption{Illustration of the model. A single-temperature quantum thermal machine where the thermal reservoir is designed employing a $\mathcal{PT}$-symmetric Hamiltonian. The position measurement process is a Gaussian operation such that the state after it is still a Gaussian state.}
\label{lastsetup1}
\end{figure}

A measurement-based quantum thermal machine with $\mathcal{PT}$-symmetric effects as the only thermal bath is explored. The cycle is illustrated in Fig. (\ref{lastsetup1}) and it is built following the notions of Ref. \cite{Yi2017}, i.e., the unitary processes are performed quasi-statically. The working substance is a single-mode quantum harmonic oscillator, described by a Hermitian Hamiltonian. The cycle is structured with two unitary and two non-unitary processes. In the unitary process, (the first and third strokes) changes in the frequency $\omega(t)$ of the working substance are implemented quasi-statically. On the other hand, the non-unitary processes are the position measurement process (second stroke) and the complete thermalization with a thermal bath (fourth stroke). We incorporate $\mathcal{PT}$-symmetry in the thermal bath through a collisional model, where each particle is composed of a single harmonic oscillator with Hamiltonian \cite{Santos2021, Behzadi2021}
\begin{equation}
    H^{\mathcal{PT}} = \frac{p^2}{2m} + \frac{m \omega^2 q^2}{2} + 2i\omega\epsilon p\,q.
\end{equation}

Note that the above Hamiltonian is $\mathcal{PT}$-symmetric and then the eigenvalues are real. We use the Dyson map $\eta = \text{exp}\left[\epsilon/\left(m\omega\hbar\right)p^2\right]$ to obtain the Hermitian counterpart such that 
\begin{equation}
    h = \frac{\mu^2}{2m}p^2 + \frac{m\omega^2 q^2}{2} + \hbar \omega \epsilon,
    \label{Hermtiancounterpart}
\end{equation}
with $\mu^2 \equiv \left(1 + 4\epsilon^2\right)$ and we vanish any $\mathcal{PT}$-symmetry signature by setting $\epsilon = 0$. 

In the simplest collisional model (CM)~\cite{Abah2020}, the thermal bath is modeled by a large number of non-interacting particles (ancillary systems), with each of them prepared in a thermal state. The assumption that the ancillas are non-interacting assures a Markovian dynamics of the working substance~\cite{Chiara2018, Ciccarello2017}. Furthermore, the CM implies that the working substance interacts with just one ancilla at a time interval $\delta t$, after which the ancilla $i$ is discarded and a new ancilla $i+1$ is brought to interact. By repeating this method many times and doing $\delta t \rightarrow 0$ the CM results in the Lindblad master equation~\cite{Breuer2002}
\begin{equation}
    \frac{d \rho }{dt} = -i\left[H_S, \rho\right] + \gamma \left(N+1\right) \mathcal{D}[a]\rho + \gamma N \mathcal{D}\left[a^\dagger\right]\rho,
\end{equation}
where $H_S = \hbar \omega_S \left(a^\dagger a + 1/2\right)$ is the Hamiltonian of the working substance, $\mathcal{D}\left[o\right] = o^\dagger \rho o - \left(\rho o^\dagger o + o^\dagger o \rho\right)/2$ is the Lindblad operator, and $N$ is the average number of photons associated with the thermal bath. Using Hamiltonian in Eq. (\ref{Hermtiancounterpart}) to build thermal states in which each ancilla is prepared, we can write them in the Fock basis as $\rho^{\text{th}} = \sum_n \left[N^n/\left(N + 1\right)^{N+1}\right] |n\rangle \langle n|$, with $N = \left(e^{\beta \hbar \omega \mu} - 1\right)^{-1}$, with $\beta = 1/\left(k_B T\right)$ the inverse temperature of the thermal reservoir. Note that it is possible to define an effective temperature $\beta_c^{\text{eff}} = \mu \beta_c$, similar to a squeezed thermal bath, in which tuning $\mu$ implies a larger or smaller temperature in the thermal reservoir.

The second non-unitary stroke is a position measurement process which is a Gaussian operation, such that since the state before the measurement is Gaussian, the state after it remains Gaussian. We choose the measurement operators to be $M_\alpha = \left(2\pi\sigma^2\right)^{-1/4} \text{exp}\left[-\left(x-\alpha\right)^2/\left(4\sigma^2\right)^2\right]$ \cite{Yi2017}. The whole cycle is described by the following strokes.

First stroke. The working substance is detached from the thermal reservoir and from the measurement apparatus. The frequency is quasi-statically changed from $\omega_1$ to $\omega_2$, such that there no transition between different eigenstates. During this process, the populations of the state remains constant. The work associated with this stroke is $W_1 = \text{Tr}\left[\rho_{\tau_1}\left(\omega_2\right) H\left(\omega_2\right)\right] - \text{Tr}\left[\rho_{0}\left(\omega_1\right) H\left(\omega_1\right)\right]$.

Second stroke. A measurement apparatus is attached to the working substance such that its position is measured. Using the position measurement operator $M_\alpha$ mentioned above, we measure the position $\alpha$ with precision given by $\sigma^2$. The post-measurement state is $\rho^M\left(\omega_2\right)$ and the energy exchanged with the measurement apparatus is $\langle Q_2^M \rangle = \text{Tr}\left[\rho^M\left(\omega_2\right) H\left(\omega_2\right)\right] - \text{Tr}\left[\rho_{\tau_1}\left(\omega_2\right) H\left(\omega_2\right)\right]$. This energy exchange is defined as heat because the Hamiltonian is kept fixed during this process.

Third stroke. The working substance is again detached from the measurement apparatus and the frequency changes back from $\omega_2$ to $\omega_1$. During this process, the populations of the state remains constant. The work associated to this process is $W_3 = \text{Tr}\left[\rho_{\tau_3}\left(\omega_1\right) H\left(\omega_1\right)\right] - \text{Tr}\left[\rho^M\left(\omega_2\right) H\left(\omega_2\right)\right]$.

Fourth stroke. In order to close the cycle, the working substance is brought to interact with the thermal bath for a sufficiently long time such that the asymptotic state is reached, i.e., $\rho_{\tau_4}\left(\omega_1\right) = \rho_0\left(\omega_1\right)$. The heat during this process is $\langle Q_4 \rangle = \text{Tr}\left[\rho_{\tau_4}\left(\omega_1\right) H\left(\omega_1\right)\right] - \text{Tr}\left[\rho_{\tau_3} \left(\omega_1\right)H\left(\omega_1\right)\right]$.

Noticing that all the processes involved in the quasi-static cycle are Gaussian and the initial state is also a Gaussian state, we can use the Fock basis to directly compute the net work $\langle W_{\text{net}}\rangle = \langle W_1 \rangle + \langle W_3 \rangle$ as well as the heat associated with the non-unitary processes. We get the results (more details in the Appendix)
\begin{eqnarray}
    \langle W_{\text{net}}\rangle &= &\frac{\hbar \mu}{2} \left(\omega_2 - \omega_1\right) \left(1 - \frac{1}{\left(2\pi\sigma^2\right)^{1/2}}\right) \coth\left(\frac{\beta_c\hbar \omega_1 \mu}{2}\right),\nonumber\\
    \langle Q_2^M \rangle &=& \frac{\hbar \omega_2 \mu}{2} \left(\frac{1}{\left(2\pi\sigma^2\right)^{1/2}} - 1\right)\coth\left(\frac{\beta_c\hbar \omega_1 \mu}{2}\right),\nonumber\\
    \langle Q_4 \rangle &=& \frac{\hbar \omega_1 \mu}{2} \left(1 - \frac{1}{\left(2\pi\sigma^2\right)^{1/2}}\right)\coth\left(\frac{\beta_c\hbar \omega_1 \mu}{2}\right).
    \label{thermoquantities}
\end{eqnarray}

For the cycle to operate as an engine (refrigerator), the thermodynamic quantities have to satisfy $\langle W_{\text{net}}\rangle < 0$, $\langle Q_2^M \rangle > 0$, and $\langle Q_4 \rangle <0$ ($\langle W_{\text{net}}\rangle > 0$, $\langle Q_2^M \rangle < 0$, and $\langle Q_4 \rangle > 0$). From the results in (\ref{thermoquantities}) we obtain the efficiency $\eta = - \langle W_{\text{net}}\rangle /\langle Q_2^M \rangle = 1- \omega_1/\omega_2$ for the engine regime and the coefficient of performance $\text{COP} = \langle Q_4 \rangle/\langle W_{\text{net}}\rangle = \omega_1/\left(\omega_2 - \omega_1\right)$. These results show that the proposed quantum machine operates in the Otto limit. This behavior is expected since we do not have the production of quantum coherence in the unitary processes~\cite{Camati2019}, such that the quantum friction along the cycle is zero. Besides the efficiency and coefficient of performance, the important aspect is how the quantum machine behavior depends  on the $\mathcal{PT}$-symmetry and the position measurement process composing the cycle.

\begin{figure}[H]
\centering
\includegraphics[scale=0.80]{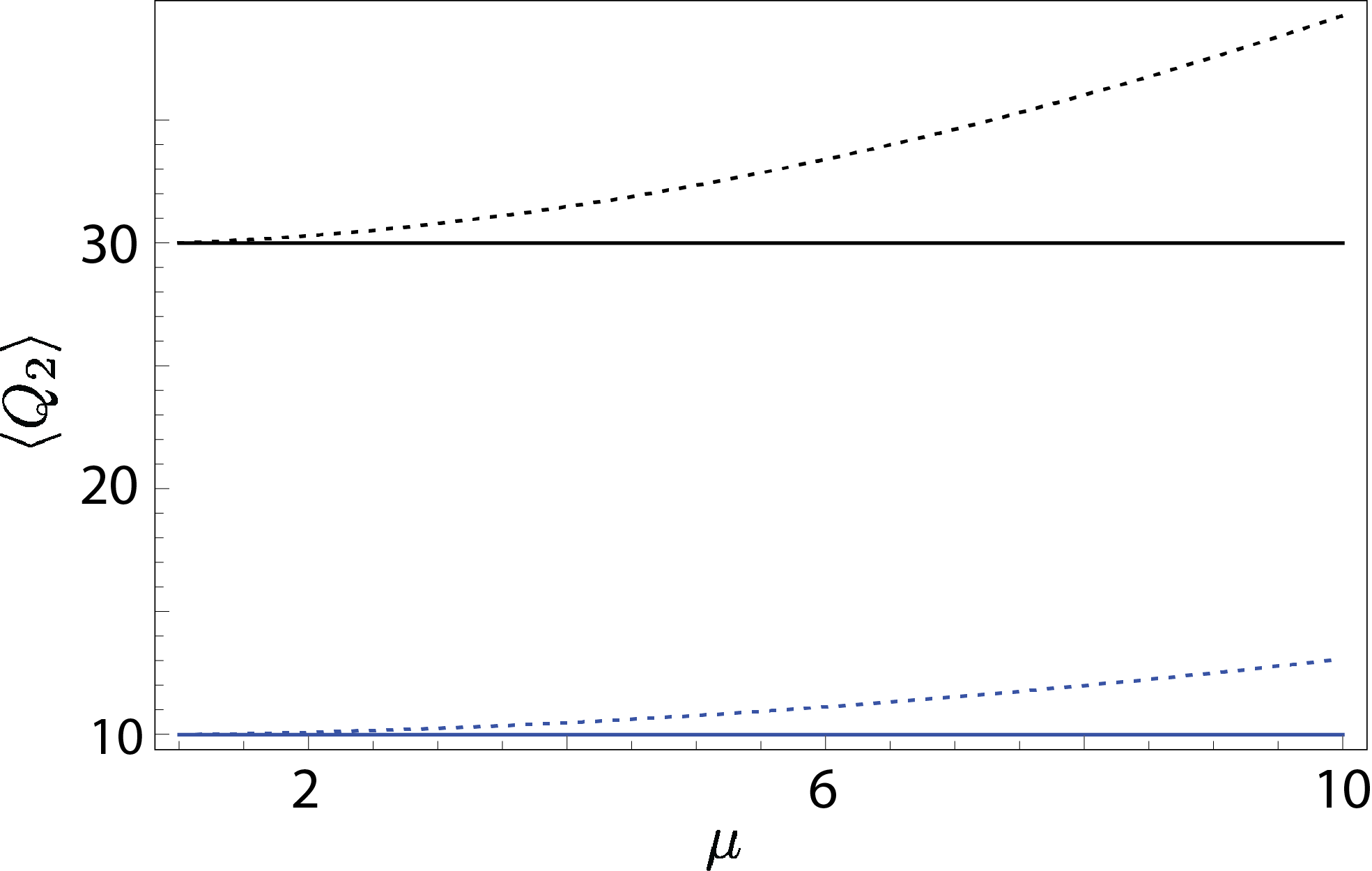}
\includegraphics[scale=0.80]{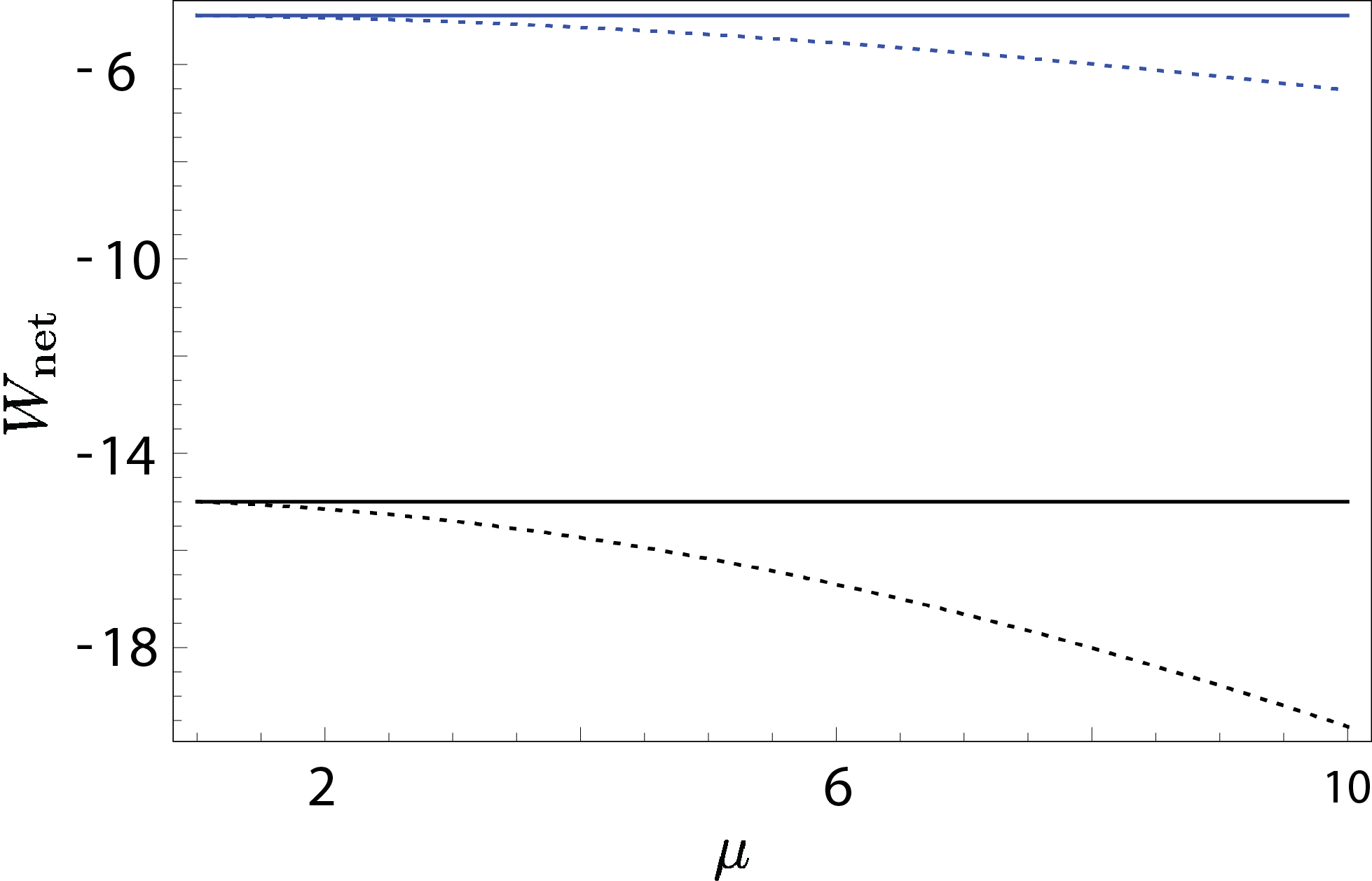}
\caption{Thermodynamics quantities. (a) Heat absorbed by the working substance during the measurement process (second stroke) and (b) net work as a function of the $\mathcal{PT}$-symmetric parameter $\mu$. Black (blue) curves are for $\sigma = 0.1$ ($\sigma = 0.2$)). Solid lines are set for $\mu = 1$, representing a cycle without $\mathcal{PT}$-symmetry in the thermal reservoir.}
\label{illustrationA}
\end{figure}

In order to better understand this behavior, Fig. (\ref{illustrationA}) depicts the heat exchanged during the measurement process and the net work as a function of the $\mathcal{PT}$-symmetric parameter $\mu$, fixing the value of $\sigma = 0.1$ ($\sigma = 0.2$) in the black (blue) curves. To compare with the scenario where we have a standard thermal reservoir, we define the solid lines for $\mu = 1$. We first note that for these values of $\sigma$, the cycle operates as an engine for any values of $\mu$. However, it can be noted that the larger the value of $\mu$, the larger the absorbed heat and net work. An interesting point is that the $\mathcal{PT}$-symmetry of the thermal bath and the measurement process are dependent on each other. This is because the $\mathcal{PT}$-symmetric property appears exactly in the part of the Hamiltonian which does not commute with the measurement operator. Thus a momentum measurement operator would not affect the thermodynamic quantities as observed in Fig. (\ref{illustrationA}).

\begin{figure}[H]
\centering
\includegraphics[scale=0.80]{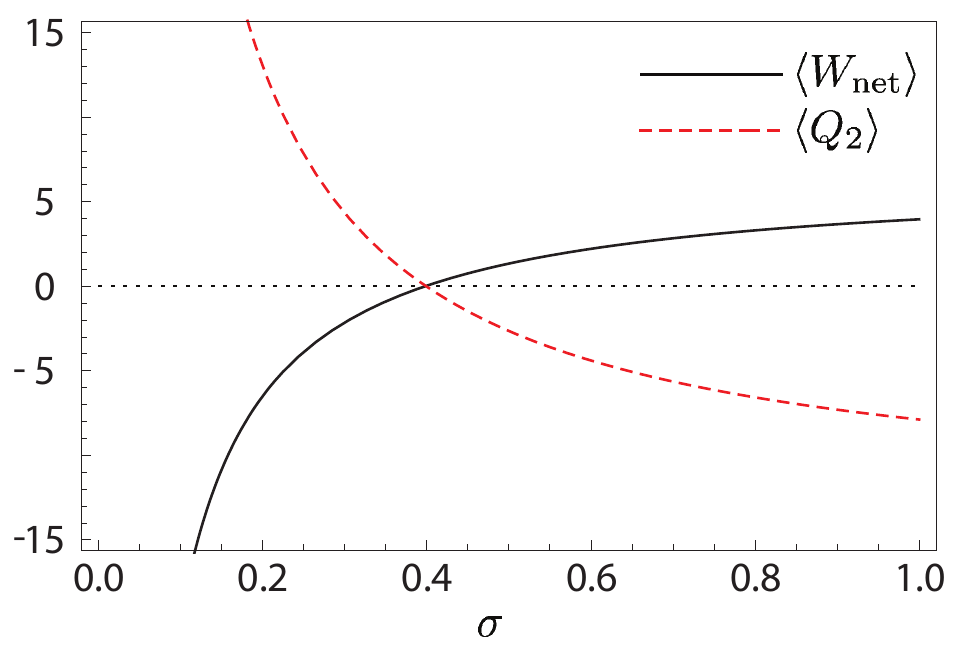}
\caption{Thermodynamics quantities. Heat absorbed/released by the working substance during the measurement process (red curve) and net work (black curve) as a function of the measurement parameter $\sigma$. We have set $\mu = 10$, $(\omega_c, \omega_h, \beta) = (1.0, 2.0, 0.2)$.}
\label{illustrationB}
\label{lastsetup}
\end{figure}

Figure (\ref{illustrationB}) shows the thermodynamic quantities as a function position measurement parameter $\sigma$ by keeping fixed $\mu$. We observe that tuning the measurement precision results in switching the quantum machine regime from refrigerator to engine. For $\sigma^2 \rightarrow 0$ we have an infinite precision (projective measurement protocol) and in this case, the machine operates as an engine. However, if the measurement process is such that the working substance is only weakly modified, then the cycle operates as a refrigerator. This is an interesting aspect of generalized measurement in quantum thermal machines which are not present in quantum cycles fueled by projective measurements. The value of $\sigma$ where the cycle regime switches is $\sigma_s = 1/\sqrt{2\pi}$. The fact that $\sigma_s$ does not depend on the temperature is because the cycle contains only one thermal bath.

%For a finite-time cycle satisfying the quantum adiabatic theorem, this result shows that if a sufficiently high control over the measurement protocol is achieved, then the cycle regime can be tuned to provide a larger value of power output (cooling rate) for the engine (refrigerator) regime.

The proposed quantum thermal machine is theoretical in nature. However, some recent advances should be highlighted in order to indicate that this cycle can be experimentally performed or simulated in the future. Optomechanical systems are interesting platforms where continuous variables can be tested. For example, $\mathcal{PT}$-symmetric dynamics have been considered in optomechanical systems~\cite{Xu2021, Longhi2016}. In particular, Ref.~\cite{Xu2016} proposes to generate a $\mathcal{PT}$-symmetric dynamics employing two-coupled optomechanical systems, which mimics exactly the collisional model we have considered in the thermal bath. On the other hand, the measurement process (second stroke), has been investigated in Ref.~\cite{Clarke} in which the position operator is measured. These advances show that an implementation of the present cycle could be possible using optomechanical systems.

\section{Conclusion}\label{conclusion}

We have considered a measurement-based quantum thermodynamic cycle in which the only thermal bath is built employing $\mathcal{PT}$-symmetric non-Hermitian Hamiltonians. With this, we introduced two non-trivial parameters in the cycle performance, the $\mathcal{PT}$-symmetry signature through the Dyson map and the measurement parameter. We stress that from our results, the relevant effect caused by the $\mathcal{PT}$-symmetry signature is to introduce an effective temperature in the thermal bath. 

The detailed analysis of the cycle showed that fixing the measurement position precision, the larger the $\mathcal{PT}$-symmetry aspect in the only thermal bath, the larger the absorbed heat and the net work. On the other hand, for a fixed value of $\mu$, varying the position measurement precision implies the conversion from the refrigerator to the engine regime. This interesting aspect sheds some light on the role played by generalized measurements in quantum thermodynamics protocols.  But from Ref. \cite{Camati2019} and from our thermodynamics quantities it is clear that this new degree of freedom does not affect the efficiency or the COP for the quasi-static case. As a future perspective, the finite-time dynamics in the unitaries processes could be investigated to study how transitions between the eigenstates modifies the $\mathcal{PT}$-symmetry and measurement protocol effects.

Finally, we have seen some advances in the experimental scenario, in particular optomechanical systems, which could be useful to realize or simulate the present quantum cycle. We hope our work can help to contribute to unveiling the role played by $\mathcal{PT}$-symmetric non-Hermitian Hamiltonians in quantum thermodynamics. 

\section*{Data availability statement}

All data that support the findings of this study are included in the article.

\section*{Acknowledgments}
J. F. G. Santos acknowledges S\~{a}o Paulo Research Grant
No. 2019/04184-5, Universidade Federal da Grande Dourados and Universidade Federal do ABC for the support.

\section*{Appendix}

We here detail the results in Eqs. (\ref{thermoquantities}). Under the collisional model assumptions in the main text, the initial state of the working substance is 
\begin{equation}
\rho_{0}=2\sinh\left[\frac{\beta\hbar\omega_{1}\mu}{2}\right]e^{-\beta\hbar\omega_{1}\mu\left(n+1/2\right)}.
\end{equation}

At $t= 0$ the Hamiltonian is $H\left(0\right) = \frac{\hbar\omega_{1}\mu}{2}\left(1+2a^{\dagger}a\right)$, such that the initial energy is 
\begin{eqnarray}
    U_{0} &=& Tr\left[\rho_{0}H\left(0\right)\right]\nonumber\\
            &=&  \frac{\omega_{1}\hbar}{2}\mu\coth\left(\frac{\beta\hbar\omega_{1}\mu}{2}\right).
\end{eqnarray}

After the first stroke the Hamiltonian is $H_{\tau_{1}}=\frac{\hbar\omega_{2}\mu}{2}\left[1+2a^{\dagger}a\right]$ and $\rho_{\tau_{1}}=\rho_{0}$, resulting in the energy 
\begin{eqnarray}
    U_{1} &=&Tr\left[\rho_{\tau_{1}}H\left(\tau_{1}\right)\right]\nonumber\\
    &=& \frac{\omega_{2}\hbar\mu}{2}\coth\left(\frac{\beta\hbar\omega_{1}\mu}{2}\right).
\end{eqnarray}

The second stroke is a position measurement process, such that the state after the measurement is
\begin{eqnarray} 
    \rho^M &=& 2A^2\int d\alpha e^{-2\left(x-\alpha\right)^{2}/\left(4\sigma^{2}\right)}\sinh\left[\frac{\beta\hbar\omega_{1}\mu}{2}\right]e^{-\beta\hbar\omega_{1}\mu\left(n+1/2\right)}\nonumber\\
    &=& \frac{\sqrt{2}}{\left(\pi\sigma^{2}\right)^{1/2}}\sinh\left[\frac{\beta\hbar\omega_{1}\mu}{2}\right]e^{-\beta\hbar\omega_{1}\mu\left(n+1/2\right)}.
\end{eqnarray}

This results in the following internal energy after the measurement
\begin{eqnarray}
    U_{2} &=& Tr\left[\rho_{\tau_{2}}H_{\tau_{2}}\right]\nonumber\\ 
    &=&\frac{1}{\left(2\pi\sigma^{2}\right)^{1/2}}\frac{\omega_{2}\hbar\mu}{2}\coth\left(\frac{\beta\hbar\omega_{1}\mu}{2}\right),
\end{eqnarray}

and the internal energy at the end of the third stroke
\begin{eqnarray}
     U_3 &=&  Tr\left[\rho_{\tau_{3}}H_{\tau_{3}}\right]\nonumber\\
     &=& \frac{1}{\left(2\pi\sigma^{2}\right)^{1/2}}\frac{\omega_{1}\hbar\mu}{2}\coth\left(\frac{\beta\hbar\omega_{1}\mu}{2}\right).
\end{eqnarray}

The thermodynamic quantities, i.e., the net work, the heat during the measurement, and the heat exchanged with the thermal reservoir are given respectively by
\begin{eqnarray}
    \langle W_{\text{net}}\rangle &=& \langle U_{1} \rangle - \langle U_{0} \rangle + \langle U_{3} \rangle - \langle U_{2} \rangle\nonumber\\
     \langle Q_2^M \rangle &=& \langle U_{2} \rangle - \langle U_{1} \rangle\nonumber\\
     \langle Q_4 \rangle &=& \langle U_{4} \rangle - \langle U_{3} \rangle.
\end{eqnarray}

and we obtain the expressions in the main text.

\end{document}